\documentclass[11pt,twoside,letterpaper]{article} 
\usepackage{natbib}
\usepackage{times,fancyhdr}
\usepackage[dvips]{graphicx}
\usepackage{url}


\makeatletter 
\def\cleardoublepage{\clearpage\if@twoside \ifodd\c@page\else%
    \hbox{}%
    \thispagestyle{empty}%
    \newpage%
    \if@twocolumn\hbox{}\newpage\fi\fi\fi} 
\makeatother

\def\figurename{Figure}
\makeatletter
\renewcommand{\fnum@figure}[1]{\figurename~\thefigure.}
\makeatother

\def\tablename{Table}
\makeatletter
\renewcommand{\fnum@table}[1]{\tablename~\thetable.}
\makeatother

\begin{document}
\title{
{\begin{flushleft}
\vskip 0.45in
{\normalsize\bfseries\textit{}}
\end{flushleft}
\vskip 0.45in
\bfseries\scshape Powerful jets from accreting black holes: evidence from the optical and infrared.}
\vskip 1in
}
\author{\bfseries\itshape David M. Russell\thanks{E-mail address: D.M.Russell@uva.nl},\\
University of Amsterdam, Netherlands\\
\\
\bfseries\itshape Rob P. Fender\\
University of Southampton, United Kingdom}
\date{}
\maketitle
\thispagestyle{empty}
\setcounter{page}{1}
\thispagestyle{fancy}
\fancyhead{}
\fancyhead[L]{In: Black Holes and Galaxy Formation \\ 
} 
\fancyhead[R]{ISBN 0000000000  \\
\copyright~2010 Nova Science Publishers, Inc.}
\fancyfoot{}
\renewcommand{\headrulewidth}{0pt}

\vspace{1in}


\vspace{.08in} 

{\abstract{
A common consequence of accretion onto black holes is the formation of powerful, relativistic jets that escape the system. In the case of supermassive black holes at the centres of galaxies this has been known for decades, but for stellar-mass black holes residing within galaxies like our own, it has taken recent advances to arrive at this conclusion. Here, a review is given of the evidence that supports the existence of jets from accreting stellar-mass black holes, from observations made at optical and infrared wavelengths. In particular it is found that on occasion, jets can dominate the emission of these systems at these wavelengths. In addition, the interactions between the jets and the surrounding matter produce optical and infrared emission on large scales via thermal and non-thermal processes. The evidence, implications and applications in the context of jet physics are discussed. It is shown that many properties of the jets can be constrained from these studies, including the total kinetic power they contain. The main conclusion is that like the supermassive black holes, the jet kinetic power of accreting stellar-mass black holes is sometimes comparable to their bolometric radiative luminosity. Future studies can test ubiquities in jet properties between objects, and attempt to unify the properties of jets from all observable accreting black holes, i.e. of all masses.
}}

\pagestyle{fancy}
\fancyhead{}
\fancyhead[EC]{David M. Russell, Rob P. Fender}
\fancyhead[EL,OR]{\thepage}
\fancyhead[OC]{Evidence of jets produced by black holes from optical and infrared observations.}
\fancyfoot{}
\renewcommand\headrulewidth{0.5pt} 

\section{Introduction}

There is currently just one way in which we can study black holes phenomenologically in the \textit{lab} -- investigating the radiation generated by their gravitational influence on matter. Here, the \textit{lab} is outer space, and the detectors are telescopes sensitive to various different wavelengths of radiation (radio, infrared, optical, ultraviolet, X-ray, $\gamma$-ray). Black holes (BHs) with masses between a few times the mass of the Sun (solar masses; $M_\odot$) and a billion $M_\odot$ attract local matter, which can convert its gravitational potential energy into heat and light as it approaches the BH. By definition, BHs are dark and do not directly produce radiation we can detect. However it is because they are massive and compact that the gravitational forces close to BHs are the most extreme in the universe; as a result the attracted matter is extremely hot and emits high-energy radiation.

The gravitational force in newtons (ignoring general relativity; this is just for purposes of demonstration) on a particle of mass $m$ kg due to a massive object of mass $M$ kg is
\begin{eqnarray*}
  F = \frac{G M m}{r^2},
\end{eqnarray*}
where $G$ is the gravitational constant $G = 6.674 \times 10^{-11}$ N m$^2$ kg$^{-2}$ and $r$ is the distance to the centre of mass of the massive object in metres.
The radius of the event horizon (beyond which no radiation or matter can escape) of a (non-spinning) BH is 
\begin{eqnarray*}
  R = \frac{2 G M}{c^2},
\end{eqnarray*}
where $c$ is the velocity of light, $3 \times 10^5$ km s$^{-1}$. From the above equations, the maximum gravitational force from a non-spinning BH (i.e. at the event horizon) is 
\begin{eqnarray*}
  F_{\rm max} \propto \frac{M}{R^2} \propto \frac{1}{M}
\end{eqnarray*}
Close to the event horizon of a $M = 10 M_{\odot}$ black hole, the gravitational force is $\sim 10^{10}$ times that near the surface of the Sun. It is the most compact objects in the universe -- the black holes and neutron stars (a neutron star is a collapsed star with a density of the order $\sim 10^{17}$ kg m$^{-3}$) that exert the largest gravitational force possible on a particle (this holds if general relativistic effects are taken into account). The gravitational potential energy of nearby matter turns into heat and radiation; the extreme gravitational conditions near BHs make them the brightest objects of high energy emission (X-rays and $\gamma$-rays) in the universe. We note that in the next decade or so it is likely that the signature of black holes will also be detected via gravitational waves \citep[via e.g. black hole binary mergers; e.g.][]{flanet98}. This will then become the second method in which we can study them observationally.

Black hole candidates in our Galaxy are thought to exist as a result of a massive star collapsing at the end of its life in a supernova explosion. Nuclear fusion in the core of the star ceases because its fuel has been depleted, causing the loss of hydrostatic equilibrium and the core collapses under its own gravitational attraction. Most Galactic BH candidates may be isolated, and accreting matter from the interstellar medium (ISM) at low, undetectable rates. The BHs that we $do$ detect are in X-ray binaries (XBs) -- binary systems in which the BH candidate (of mass $M \sim 3$ -- $20 M_{\odot}$ or more) and a (relatively normal) star are orbiting a common centre of mass. If the surface of the star lies close to the point between the two objects in which the gravitational force from both are equal (the L1 point), the outer layers leak as a stream towards the BH. The net angular momentum of the gas causes it to form a disc around the BH as the matter spirals inwards. The BH is hence being fed, via this accretion disc, by a semi-regular supply of gas from the companion star. As the matter moves inwards its temperature increases; the highest energy radiation is emitted from the hottest gas close to the BH. A visual illustration of an accreting black hole candidate X-ray binary (BHXB) is given here: \url{http://www.phys.lsu.edu/~rih/binsim/binarysim_annotate.jpg}

\subsection{Powerful jets from supermassive black holes}

It is somewhat counterintuitive that it is in these extreme gravitational fields close to BHs that the highest velocity \emph{outflows} in the universe are formed. One can use the simple argument however that the matter must lose a large amount of its potential energy and angular momentum if it is to reach the event horizon, and outflows are one way of removing this problem by channelling some of the energy and angular momentum away from the BH; some matter then never reaches the event horizon. Jets of matter and energy travelling close to the speed of light were first discovered originating from the vicinity of supermassive ($M \sim 10^5$ -- $10^9 M_{\odot}$) BHs in active galactic nuclei (AGN) at the centres of galaxies. These are the most massive BHs known in the universe, and their powerful, relativistic jets, and the interactions of those jets with local matter, have been studied for decades \citep[e.g.][]{baadmi54,fanari74}. Hotspots indicate where the jets impact with the intracluster gas, and lobes of synchrotron-emitting plasma are inflated which flow back from the hotspot, both of which are often detected at radio frequencies. AGN and their jets are now realised to play an important role in the evolution of matter in the universe, affecting to some extent the rate of galaxy growth in the early universe \citep[e.g.][]{efst92,churet02,mcnaet07,schaet09}.

There are at least three fundamental questions that remain unanswered regarding AGN jets:

\begin{itemize}
\item What is the process of particle acceleration?
\item What is their composition?
\item How much influence do they have on shaping large-scale structure in the universe?
\end{itemize}

Difficulties arise in answering these questions because various models predict similar observational properties. For example, the composition of the jets and radio lobes, which are not charged, could be either baryonic (electrons and protons) or leptonic (electrons and positrons), both of which result in synchrotron emission which is observed. Similarly, the energy required to accelerate the jets to relativistic velocities could be tapped either by extraction of the spin energy of the BH, or from the accretion energy in the accretion disc \citep{blanzn77,blanpa82,meie01}; these have not yet been distinguished by observations. The power contained in AGN jets \emph{is} known, at least to some degree. It is possible to use (successfully in some cases) the jet hotspots and lobes to infer the power input into them, i.e. the energy contained in the jets averaged over time \citep{burb59}.

\subsection{Powerful jets from stellar-mass black holes}

It is only in recent years that it has become apparent that accretion onto stellar-mass BH candidates (and neutron stars) in X-ray binaries commonly produce relativistic, synchrotron-emitting jets in analogy with those of AGN \citep[e.g.][]{miraet92,falcet04,fend06,miglfe06}. The name `microquasar' was coined for X-ray binaries with jets due to this analogy. The jets here are found to be associated with specific X-ray regimes \citep{fendet04}. When the X-ray spectrum is hard, a steady jet exists, the radio luminosity of which correlates with the X-ray luminosity \citep{gallet03,gallet06}. This correlation extrapolates successfully to the radio and X-ray luminosities of AGN when introducing a mass term which is consistent with that predicted by scalings in black hole physics \citep{merlet03,falcet04}, demonstrating that a single jet formation process alone likely governs the physics of all BHs. When BHXBs make transitions to softer X-ray states, bright discrete ejecta are observed and often resolved at large distances from the BH \citep[e.g.][]{tinget95,gallet04}. The core jet emission is then quenched, and only returns when the X-ray spectrum again hardens \citep{fendet04}. XBs vary in luminosity (and hence mass accretion rate) by orders of magnitude on timescales accessible to us -- typically over several weeks or months. By comparison, AGN take millions of years to perform these `outburst' cycles. A recent step forward in our understanding was made when it was proposed that the different X-ray regimes and jet behaviour of BHXBs can be directly linked to the different \emph{classes} of AGN \citep{kordje06}. Short term variability probes the inner regions of the accretion flow close to the BH, in the vicinity where the jets are formed. These timescales scale with the mass of the BH; for BHXBs it is milliseconds to minutes, whereas for AGN it can be from hours to years. Minute-to-hour long observations can hence probe these timescales in BHXBs, but dedicated long-term campaigns are required for the same study of AGN.

It is uncertain how powerful XB jets are. Few sites of interaction between the jets of XBs and the ISM have as yet been identified but there are a growing number of examples (see Sections 2.2 and 2.3). These interaction sites can be used to infer the time-averaged power of XB jets \citep{kaiset04,gallet05}. Estimates of the jet power based on the core jet luminosity have large errors because both the total spectrum of the jet and its radiative efficiency are poorly constrained \citep[however see][for a recent advancement of the latter]{kordet06a}. Estimates of the jet power are dominated by the highest energy electron distributions in the jet, which generate the highest energy photons, but isolation of the jet spectrum can be hampered by other emitting components such as the accretion disc at these higher energies (see Section 2.1). Measuring as accurately as possible the jet power ($P_{\rm J}$) in different sources and at different luminosities is crucial in understanding the process of jet formation and the overall physics of accretion and the matter and energy BHs input into the ISM.

\begin{figure}
\centering
\includegraphics[width=21cm,angle=270]{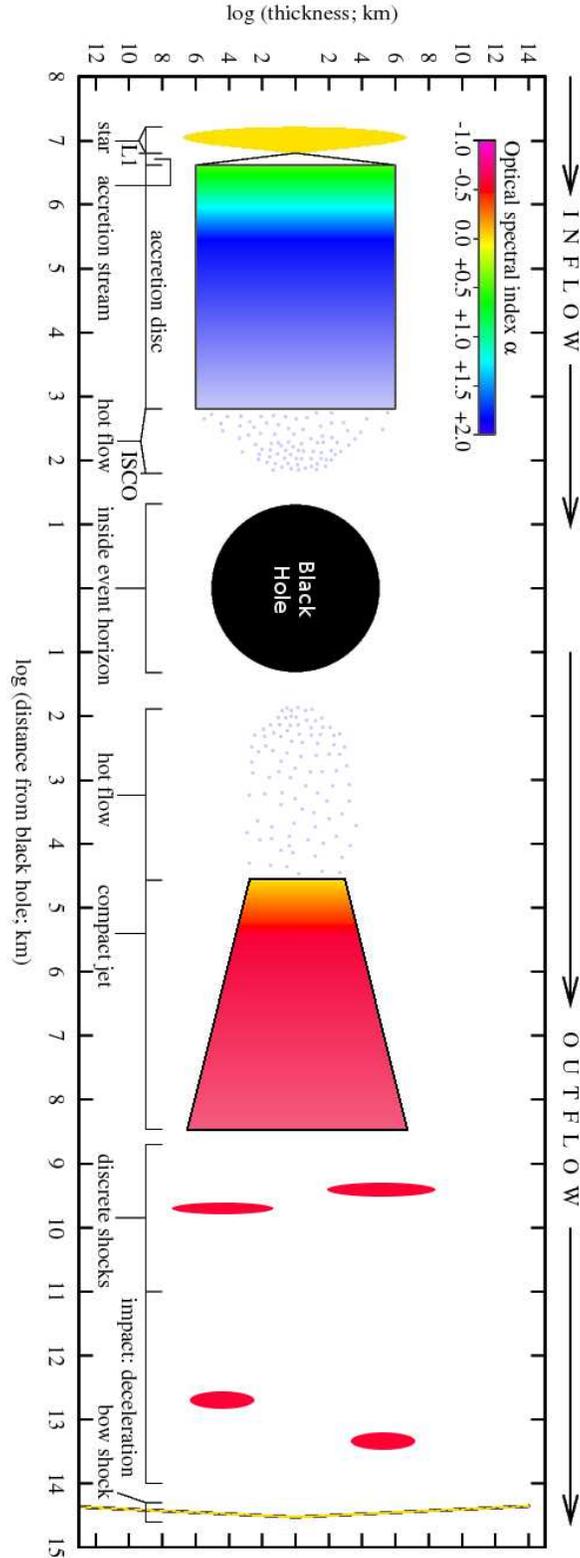}
\caption{A schematic demonstrating the approximate dimensions and optical colours of the components of a low-mass BHXB (see text for details).
}
\label{fig:fig1}
\end{figure}

It may be that stellar-mass black holes in BHXBs hold the key to understanding AGN and their jets. If the BHXBs are indeed simply scaled-down versions of AGN, with similar physical structure and behaviour, the three unanswered questions above regarding AGN jets may be answered via studies of BHXBs because they vary on accessible timescales, and their spectra are not complicated by light from the host galaxies or affected (as much) by orientation and obscuration. In addition, it is possible to model the composition of BHXB jets using information from both the compact jet and jet--ISM impact sites \citep{hein06}.

\subsection{optical and infrared-emitting components of X-ray binaries}

Many physical components and mechanisms are responsible for optical/infrared (OIR) emission from BHXBs and neutron star X-ray binaries. Fig. 1 is a schematic which illustrates the different components of the accretion inflow and outflow (jets), their approximate physical sizes and optical colours, for a BHXB with a low-mass companion star. The colours correspond to approximate intrinsic optical ($\sim$ V-band) spectral indices. The distances from the centre of the black hole are in logarithmic scale. The dimensions of the companion star, accretion stream and accretion disc are based on those quoted in \cite{obriet02} for the BHXB GRO J1655--40. The companion star of GRO J1655--40 has an approximately flat spectral index at V-band \citep[$\alpha = 0$;][]{miglet07} so it appears yellow. The accretion disc, which produces a multi-temperature black body spectrum is hotter at smaller radii, and the optical spectral index steepens in this case from $\alpha \sim +0.5$ at its outer edge to probably $\alpha \sim +2.0$ in the inner regions \citep[based on observed values during outbursts but these change with disc temperature; e.g.][]{hyne05}. Although hotter, the emitting region is smaller in the inner regions of the disc, and the bulk of the optical emission from the disc originates in the outer regions. The inner edge of the accretion disc is at a radius $\sim 10^{-4}$ times that of the L1 radius \citep[][L1 is the point at which the gravitational force from the BH and the star is equal; material from the star is no longer gravitationally bound to the star if it crosses this point towards the BH]{obriet02}. The hot, optically thin flow (which may also exist above and below the inner part of the disc) ends at the innermost stable circular orbit (ISCO), which is at three times the radius of the event horizon. The radius of the event horizon here is 21 km (we assume a non-spinning black hole of mass $M = 7 M_{\odot}$).

The outflow illustrated in Fig. 1 comprises a compact, continuously replenished jet producing synchrotron emission, internal shocks further down the jet, deceleration of jet knots as they impact the ISM, and a radiative bow shock energized by the jet. All components are based on observed, resolved phenomena and are expected to produce OIR emission. We provide a comprehensive review and discussion of the OIR signatures of jets in Section 2.

The compact, conical jet, which is accelerated to relativistic velocities close to the BH, produces synchrotron radiation from electron distributions with energies decreasing with distance from the BH. The optical depth also decreases as the ejected material expands as it flows down the jet. The result is a series of overlapping, self-absorbed synchrotron spectra which form an approximately flat ($\alpha \sim 0$) radio-to-infrared spectrum, where the synchrotron spectrum from each electron energy distribution peaks at a frequency which is approximately inversely proportional to the distance from the jet origin \citep{blanko79,falcbi96}. \cite{stiret01} resolved the compact steady jet of Cygnus X--1 at 8.4 GHz. At this radio frequency its extent was 15 milli-arcseconds (mas) $\approx 3 \times 10^8$ km adopting a distance to the BHXB of 2.1 kpc \citep{masset95}. Assuming the highest energy electron distributions in the inner regions of the jet produce a synchrotron spectrum that peaks in the near-infrared (NIR; at $\nu \sim 10^{14.5}$ Hz, as is implied by some observations; see Section 2.1) the summed spectrum will break to an optically thin one, with $\alpha \sim -0.7$ blueward of this `jet break'. The corresponding radius of the innermost region of the synchrotron-emitting compact jet is then $\sim 4 \times 10^4$ km. This is also consistent with fast timing observations in the optical, in which variability on timescales corresponding to emitting regions as small as $\sim 2 \times 10^4$ km have been shown to most likely originate in the inner regions of the jet \citep[e.g.][see Section 2.1]{kanbet01,hyneet03b}. The opening angle of the radio jet of Cyg X--1 was $< 2^{\circ}$ \citep{stiret01}; here we assume $2^{\circ}$. This is consistent with measurements from other BHXBs \citep{millet06} but much larger than implied by some models \citep[e.g.][]{kais06}. The outer region of the jet that produces the 8.4 GHz radio emission would then be wider than the diameter of the accretion disc (assuming the opening angle is constant along the jet). The compact jet cannot currently be resolved at OIR frequencies. The inner regions closest to the BH ($< 10^5$ km away say) dominate the OIR emission from the compact jet; at a distance of 1 kpc (the closest known BHXBs are at this distance), this corresponds to an angular size of the order of micro-arcseconds, which is not resolvable with current detection limits \citep[although sophisticated interferometric methods may achieve this in the future;][]{mark08}. The radio emission from the compact jet originates in a region much larger than this, but this spectral component is orders of magnitude fainter at OIR frequencies and cannot be detected.

BHXB jets \emph{have} however been resolved at radio and X-ray frequencies, and can be seen to move at relativistic speeds, at distances larger than $\sim 10^9$ km from the BH (see Section 2.2). The emission process is thought to be optically thin synchrotron which peaks at low frequencies, and could either be energized by shocks between discrete ejecta travelling at different velocities, or by impacts with the ISM, which decelerate the ejecta. In Fig. 1 discrete shocks are shown at distances $10^{9-10}$ km \citep[which have been seen in the radio in Sco X--1, actually a neutron star XB;][]{fomaet01} and impacts with the ISM are shown at $\sim 10^{13}$ km \citep[here observed from the BHXB XTE J1550--564;][]{corbet02}. There is just one tentative claim so far of this resolved synchrotron emission detected at OIR frequencies: from GRS 1915+105 \citep{samset96}. For the above case of XTE J1550--564 the authors point out that the synchrotron spectrum from radio to X-ray is just below the detection limits of optical data taken with the Very Large Telescope.

It has been shown that the impact of the jets of BHXBs with the ISM inflate lobes of radio plasma, like those seen in FR II radio galaxies powered by the jets of AGN \citep[e.g.][]{kaiset04}. Ahead of these lobes there is a radiative shock wave, which produces both thermal continuum and line emission from the recombination of ionized atoms. Optical line emission and thermal radio emission from a bow shock powered by a BHXB jet has recently been detected -- at a distance $\sim 3 \times 10^{14}$ km ($\sim 10$ parsecs) from Cygnus X--1, with a diameter approximately half that distance. This is the furthest and largest structure illustrated in Fig. 1 (see Section 2.3 for discussion of these structures).

\section{OIR identifications of jets from black hole X-ray binaries}

Traditionally, jets have been identified in BHXBs from the radio regime.  This is firstly because they are sometimes resolved, which is a very strong argument for a jet, and secondly because no other known emitting components can account for the radio fluxes and spectra \cite[see][for a review]{fend06}. There is now strong empirical evidence for the spectrum of the compact jet to extend to, and be detected at, OIR frequencies and higher. Although these detections are complicated by other spectral components (which is usually not the case at radio frequencies), many methods can be adopted to successfully disentangle the components and isolate the jet emission. It is in some cases \emph{easier} to detect BHXB jets in the OIR with current facilities. Signatures of OIR jet emission include its spectrum, timing properties and polarimetry properties, all of which differ from other OIR--emitting components such as the accretion disc. In Section 2.1 we review these signatures and discuss the observations of OIR jets in the literature. We stress that although our review focusses on the jet emission, the process that \emph{dominates} the optical (and in some cases the NIR too) emission of most transient BHXBs is X-ray reprocessing on the surface of the accretion disc. For BHXBs with high-mass stars, and for BHXBs at low accretion rates in quiescence, the companion star usually dominates the OIR light.

As was discussed in Section 1.3, large-scale, \emph{resolved} OIR emission can be produced directly and indirectly by the jets of BHXBs. In Sections 2.2 and 2.3 we summarise the known resolved OIR jets and jet--ISM interaction sites. In Section 3 we briefly discuss how these observations (OIR compact jets, resolved jets and jet--ISM interactions) can be used to infer the jet properties, and their implications for jets and accretion onto supermassive BHs in AGN. A summary is provided in Section 4, and we comment on future work that will help to answer the open fundamental questions regarding jets and accretion.

\subsection{The compact jet}

Models of synchrotron emission from compact jets of BHXBs can reproduce the approximately flat ($\alpha \sim 0$), optically thick radio spectrum of AGN and BHXBs, which breaks to one which is optically thin (the aforementioned `jet break') at some higher frequency \citep[e.g.][see Fig. 2 for an example spectrum]{blanko79,market01,market05,kais06,jamiet08}. For BHXBs the jet break probably lies within the OIR regime (see below). The jet break frequency ($\nu _{\rm b}$) depends on the distance of the jet launch region from the BH, which is related to the BH mass, and may also have a weak dependence on mass accretion rate and therefore luminosity \citep{heinsu03,nowaet05}. In addition to the spectrum, the OIR-emitting jet has been identified via polarimetry and correlations with other wavebands over short timescales ($\sim$ seconds and less; fast timing) and long timescales ($\sim$ weeks to years; `slow timing', i.e. over many orders of magnitude of mass accretion rate).
\newline\newline
\textbf{A synchrotron OIR spectrum:}

A flat ($\alpha \sim 0$) radio-to-OIR spectrum implied from radio and OIR data alone is consistent with (but not direct evidence for) a significant jet contribution to the OIR, since radio emission in BHXBs originates in the jet. \cite{fend01} pointed out that this is the case in the spectral energy distributions (SEDs) of the BHXBs V404 Cyg and GRS 1915+105. In fact, \cite{hanet92} showed that the radio and optical light curves of the decay of V404 Cyg follow similar shapes (three separate power-law decay slopes) while remaining proportional in flux ($F_{\rm \nu,radio}\propto F_{\rm \nu,optical}$). Similar flat radio--OIR SEDs are seen for GS 1354--64 (BW Cir) and GRO J0422+32 in the hard state \citep[i.e. when the jet is on;][]{brocet01,brocet04} and for GRS 1915+105 when its flickering radio jet is on \citep[e.g.][]{fendpo00,kleiet02}. A global correlation between the OIR and X-ray luminosities of 15 BHXBs in the hard state \citep[][see below]{russet06} implies that all BHXBs approximately follow the correlation $L_{\rm \nu,radio} \approx L_{\rm \nu,optical}$. Nevertheless, this correlation alone is not a \emph{diagnostic} of a jet origin to the OIR luminosity but is consistent with this interpretation.

\begin{figure}
\centering
\includegraphics[width=9cm,angle=0]{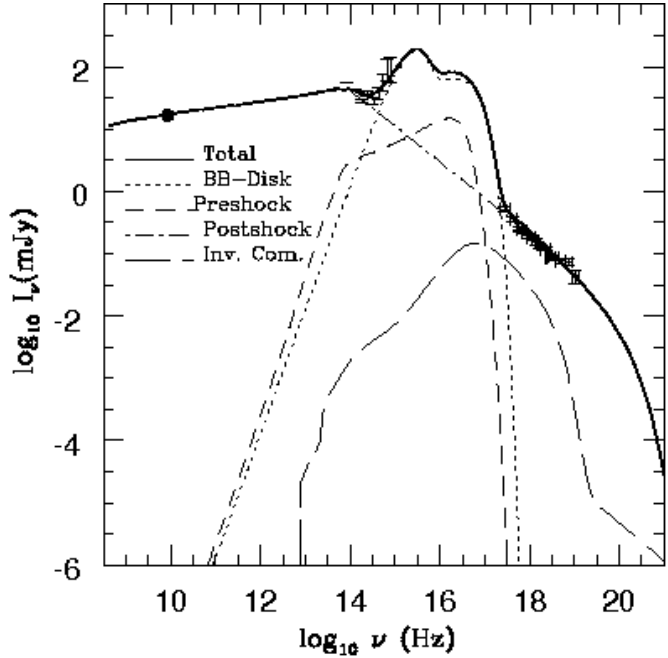}
\caption{An example of a jet model that successfully reproduces the radio, OIR and X-ray SED of a BHXB (GX 339--4 in this case); from \cite{market03}. The NIR is modelled as optically thin synchrotron emission from the jet.
}
\label{fig:fig2}
\end{figure}

More convincing is a measure of the intrinsic OIR spectral index. \cite{corbfe02} found two components of OIR emission in an SED of GX 339--4 -- one red ($\alpha < 0$) and one blue ($\alpha > 0$). The red component dominates the NIR and is consistent with optically thin synchrotron emission from the jet, for which we expect $\alpha \sim -0.7 \pm 0.2$; this value depends on the electron energy distribution. A $\sim$ flat radio--NIR spectrum was also evident in the SED. The red NIR component was seen additionally in separate outbursts of GX 339--4 \citep{homaet05}, and \cite{motcet85} measured the power-law optical---NIR spectral index to be $\alpha \sim -0.58$, fairly typical of optically thin synchrotron emission. The SEDs of XTE J1118+480 \citep{hyneet00,fendet01,chatet03a,hyneet03b} and 4U 1543--47 \citep{buxtet04,kaleet05} have similar red spectral indices in the hard state, and a NIR-excess was apparent above the spectrum of the accretion disc in XTE J1859+226 \citep{hyneet02}. SEDs of LMC X--3 also reveal an anomalous NIR-excess \citep{pine84,trevet88}, which may be the first evidence for a jet from a BHXB in the Magellanic Clouds (see also Section 2.3 for an other). OIR SEDs of V404 Cyg, GS 1354--64, GS 2000+25, XTE J1550--564 and GRO J1655--40 are sometimes red \citep[$\alpha < 0$;][]{brocet04,russet06} which is indicative of either a cold disc (possible for sources at low luminosity) or a synchrotron jet. It is possible to use OIR colour-magnitude diagrams to successfully separate thermal disc from non-thermal jet emission since the irradiated disc produces a predictable relation between OIR colour and magnitude as the temperature of the outer disc (and luminosity) changes, whereas the OIR colour of optically thin synchrotron emission is not expected to change with disc temperature or luminosity \citep{maitba08,russet08}. Colours redder than expected for a disc (deviations from a colour--magnitude relation for a heated black body) were seen in five BHXBs; the redder colours of some were argued to be due to the jets.

One claim of the jet break itself observed in the NIR SED of GX 339--4 \citep{corbfe02} is based on three data points; follow-up near simultaneous photometry is required to confirm this (the source is known to have large amplitude variability on short timescales). However the jet break may have been detected in GRO J0422+32 \citep[a self-absorbed synchrotron model fits the double-power-law optical--UV spectrum better than an accretion disc model;][]{hyneha99,shraet94} and V404 Cyg \citep{brocet04,russet06} but these are speculative.

A further source of evidence for OIR jets is successful modelling of broadband spectra, incorporating self-consistent physical constituents and emission processes. This approach has led to further confirmation of OIR jets dominating the SEDs of XTE J1118+480 \citep{market01} and GX 339--4 \citep[][an example model from this paper is shown here in Fig. 2]{market03}. For GRO J1655--40, and in quiescence for A0620--00, XTE J1118+480 and V404 Cyg, the jet dominates only at lower frequencies -- in the mid-infrared -- but makes a low-level contribution to the OIR \citep{miglet07,gallet07}.
\newline\newline
\textbf{OIR--X-ray correlation in the hard state:}

\cite{homaet05} present a correlation between the quasi-simultaneous NIR and X-ray fluxes of GX 339--4 which span two orders of magnitude in $F_{\rm X}$ during its hard state. The correlation, $F_{\rm NIR} \propto F_{\rm X}^{0.5}$ is synonymous to the hard state radio--X-ray correlation $F_{\rm radio} \propto F_{\rm X}^{0.6}$ \citep{gallet06}, suggesting again that the radio and NIR emission origins may be identical. In \cite{russet06,russet07b} it was found that the X-ray and OIR hard state data (B-band to K-band) of 15 BHXBs are consistent with a global correlation, $L_{\rm OIR} \propto L_{\rm X}^{0.6}$ which holds over eight orders of magnitude in $L_{\rm X}$. Individual BHXBs do not necessarily display this correlation within their own data sets, but all the data lie close to the global correlation. However, a correlation with a similar ($L_{\rm OIR} \propto L_{\rm X}^{0.5}$) slope is expected \citep{vanpet94} if the OIR is reprocessed emission from X-rays illuminating the accretion disc. From the observed BHXB correlation alone, the two cannot be confidently separated \citep{russet06}.
\newline\newline
\textbf{An optical/infrared flux drop when the radio jet is quenched:}

A strong indication of an OIR jet is a dramatic change of flux during an X-ray state transition. The compact radio jet is quenched in the soft state, dropping below detection limits \citep{gallet03}. \cite{homaet05} saw a drop of a factor 18 (3 magnitudes) in the NIR flux of GX 339--4 when the source entered the soft state. In that time the X-ray flux also decreased but only by a factor $\sim 2$. There are a number of examples of OIR rises/drops seen during transitions in/out of the hard state; in Fig. 3 we plot the light curves and amplitudes of this drop as a function of frequency. It appears that the change in flux is in general stronger in the NIR than in the optical, likely due to the accretion disc dominating moreso at higher frequencies because its spectrum is bluer than that of the jet. In at least one case (A0620--00) there is an X-ray rise of similar amplitude to the OIR rise -- for this source X-ray reprocessing cannot be ruled out. Moreover, \cite{russet06,russet07b} found that the NIR (but not the optical) luminosity of all six BHXBs in their study (the first three sources in the above Fig. 3, plus GRO J1655--40, XTE J1720--318 and XTE J1859+226) is weaker in the soft state than in the hard state at the same X-ray luminosity, by a factor $\sim 10$ (this cannot be accounted for by the change in X-ray spectrum). The NIR soft state data of these six BHXBs lie below the global hard state OIR--X-ray correlation. In a related result, a NIR-to-UV jet spectrum in GRS 1915+105 is implied by a drop in the NIR emission line Br-$\gamma$ when the jet continuum drops, which suggests the line is radiatively pumped by high energy photons from the jet \citep{eikeet98a}. However, this seems not to be the case for the H$\alpha$ line in BHXBs since no apparent change is seen in the equivalent width of this line after state transitions \citep{fendet09}.
\newline\newline
\textbf{Fast timing signatures:}

\begin{figure}
\centering
\includegraphics[width=6cm,angle=270]{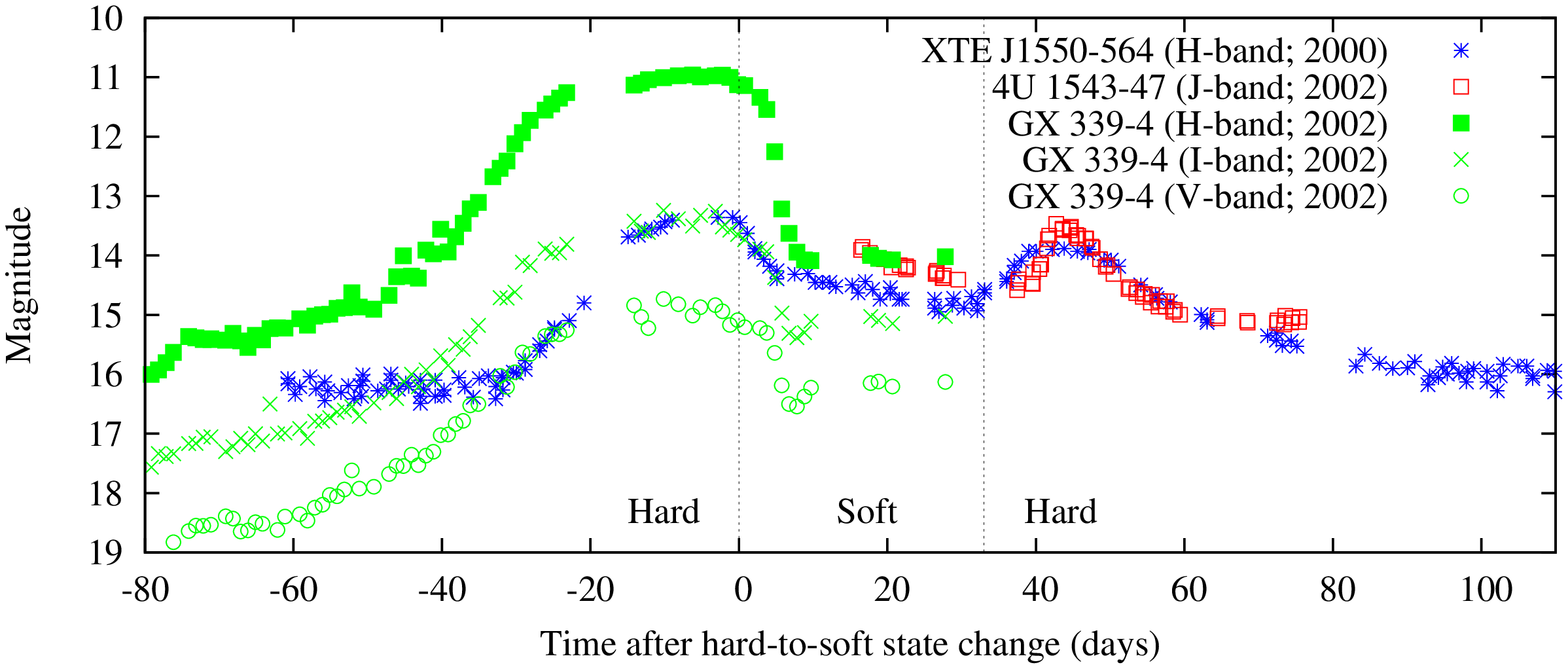}\\
\includegraphics[width=8cm,angle=270]{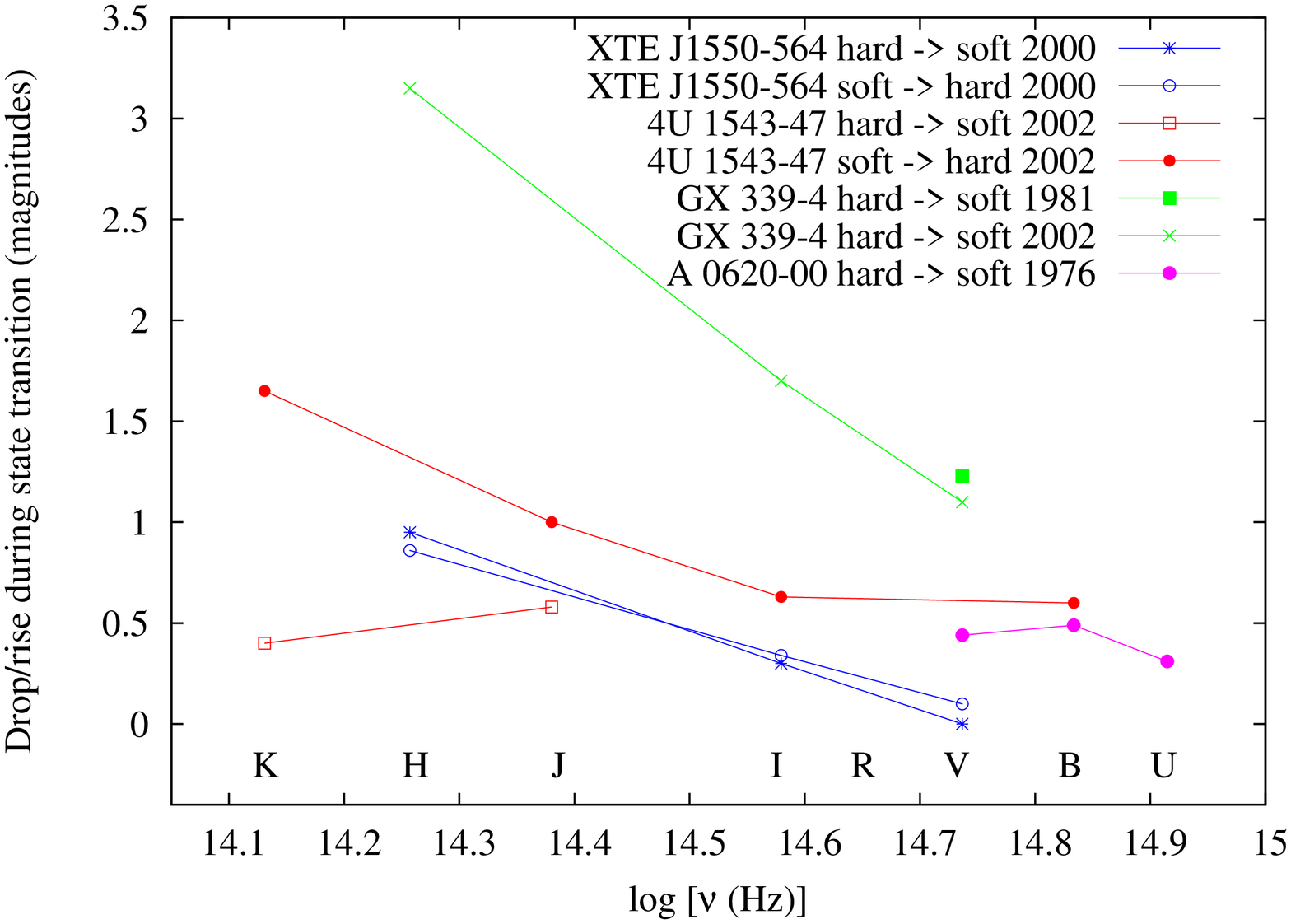}
\caption{Changes in OIR magnitudes of BHXBs over X-ray state transitions. Top panel: OIR light curves of three BHXBs. Bottom panel: Amplitude of the drop/rise during transition, as a function of frequency. The bandpasses (filters) U (ultraviolet) to K (NIR) are indicated. In all cases, the change of flux is a rise when in transition to the hard state and a drop when the source leaves the hard state (into a softer state). The data are from \cite{motcet85}, \cite{kuul98}, \cite{jainet01}, \cite{buxtet04} and \cite{homaet05}. Errors are typically less than $\sim 0.1$ mag.
}
\label{fig:fig3}
\end{figure}

Rapid (on timescales down to milliseconds), high-amplitude optical variability has been reported from BHXBs since the 1980s \citep{motcet82}. \cite{fabiet82} argued that clouds of plasma with surface brightness temperatures $\sim 5 \times 10^9$ K emitting cyclotron radiation can account for the rapid optical flares of GX 339--4. \cite{motcet83} constructed a cross-correlation function (CCF) of simultaneous fast optical and X-ray timing data of GX 339--4 during its hard state and discovered a complex behaviour in which (a) a weak positive optical response to X-ray flickering peaked after tens of seconds, and (b) an optical--X-ray anti-correlation exists whereby optical variations precede by a few seconds those in X-ray. This `precognition dip' was also later found in the CCF of XTE J1118+480 in the hard state, in addition to a strong positive optical lag \citep{kanbet01}. The optical variability is less smeared than the X-ray, so the optical origin cannot be X-ray reprocessing on the disc. A further recent study of GX 339--4 in the hard state \citep{gandet08} showed an optical--X-ray CCF with a weak precognition dip and a large but narrow positive optical response.

Two separate lines of reasoning lead to the conclusion that the origin of the fast optical flickering in XTE J1118+480 (and presumably GX 339--4) is the inner regions of the jet. \cite{hyneet03b} studied the NIR, optical, UV and X-ray variability properties and by isolating the variable component, found the variability to have a single power-law SED from NIR to X-ray, with a spectral index $\alpha = -0.6$, implying an optically thin synchrotron origin. The variability is also more lagged at lower frequencies, which is consistent with synchrotron bubbles which become optically thin at lower frequencies as they expand, and not consistent with either X-ray reprocessing or advection-dominated accretion flow models. \cite{hyneet06} found a similar spectral index, $\alpha = -0.8$ for the variable component of the NIR emission of XTE J1118+480 during a separate outburst. An alternative approach \citep{malzet04} used a time-dependent model to approximately reproduce the CCF (including the precognition dip) and power spectrum of XTE J1118+480. The model envisages a jet--disc coupling in which accretion (magnetic) energy can be channelled into either the jet power or the X-ray `corona' in a correlated way, the jet producing the optical emission.

Further clues come from optical variability studies of other BHXBs in outburst and quiescence. Brightness temperature arguments were used to rule out X-ray reprocessing in V4641 Sgr and GRO J0422+32 during hard state outbursts, favouring a non-thermal optical variability origin \citep{bartet94,uemuet02}. It was demonstrated that short ($< 100$ sec) flares from V4641 Sgr are likely synchrotron in nature, and longer ($> 1000$ sec) flares are thermal \citep[from their SEDs;][]{uemuet04a,uemuet04b}. An optical--X-ray CCF of GRO J1655--40 during a \emph{soft} state \citep{hyneet98} displays a positive optical lag on light-travel timescales, with no precognition dip. This favours a reprocessing response on the disc surface and further supports the idea that the jet is responsible for the precognition dip. However, the CCF of SWIFT J1753.5--0127 (a BH or neutron star primary is debated in this source) displays a deep, wide precognition dip and weak positive response, at a time in which the contribution of the jet to the OIR SED appears to be minimal \citep{duraet08,duraet09}.

Rather different variability results come from the BHXB GRS 1915+105 in outburst. This source is unique with respect to its X-ray timing and radio jets -- it does not remain in the canonical hard or soft X-ray states for weeks to months like other BHXB transients, but instead displays complex states at high accretion rates accompanied by correlated (sometimes resolved) radio jet ejections. Discrete, quasi-periodic NIR, mm and radio flares repeat intermittently on timescales of 30--45 minutes and are correlated with more rapid X-ray variations \citep[e.g.][]{fendet97,uedaet02,rothet05}. The NIR flares cannot be reprocessed X-rays, and instead are shown to be the precursors of the mm and radio flares -- likely single jet 
ejection events which propagate away from the BHXB, peaking at lower frequencies at larger distances \citep{miraet98,fendpo98,eikeet98b,fendpo00}. A similar process may be evident in SS 433 (this system is most likely a BHXB but may be a neutron star XB). \cite{goraet98} report a variable `red' component in the optical SED -- these optical flares lead radio flares from the SS 433 jets by several hours in time. \cite{chaket05} also claim a NIR--radio correlation on longer timescales -- a broad dip in the NIR and radio light curves is consistent with the NIR leading the radio by two days.

Numerous fast timing studies of BHXBs have been performed during periods of quiescence, but only one (to our knowledge) makes use of simultaneous X-ray coverage, because most systems are too faint for current X-ray telescopes at these luminosities. \cite{hyneet04} present a CCF between X-ray flux and H$\alpha$ optical emission line flux, which shows a positive correlation consistent with zero lag. This result combined with the observation of a double peaked H$\alpha$ line profile implies line variability in the disc powered by X-ray irradiation. Fast optical continuum variations and flaring were reported in quiescence for at least six BHXBs \citep[A0620--00, GU Mus, MM Vel, V404 Cyg, GRO J0422+32, GS 2000+25;][]{hyneet03a,zuriet03,shahet03,shahet04}. The rapidity implies (to the authors) localised flares which may occur when magnetic loops reconnect on the disc, although the power density spectrum, when measured, can resemble that of an X-ray power density spectrum of a BHXB in the hard state, in which the X-ray variability comes from the inner regions of the accretion flow. In some cases, the spectral index of the optical flares themselves have been measured and are quite steep: $\alpha = -1.4$ for A0620--00 and $\alpha = -2.0$ for V404 Cyg, which are consistent with an optically thin gas at a temperature (8--14)$\times 10^3$ K. According to the quiescent broadband SEDs of these two BHXBs \citep{gallet07}, the jet likely contributes $< 10$\% (most likely $\sim 1$\%) of the optical emission at these low luminosities, so the jets are unlikely to be responsible for the flares. However, the flares are short-lived and could be missed in the broadband SEDs, which are based on mean fluxes over longer timescales. If the flares originate in the jets, the optical--radio jet spectrum would have to be inverted; $\alpha \sim > +0.3$ for the jets to be as faint as observed in the radio regime.
\newline\newline
\textbf{Polarization signatures:}

\cite{fabiet82} predicted that if the rapid optical flaring of GX 339--4 \citep{motcet82} has a cyclotron origin, the flares should be polarized. Optically thin synchrotron emission can be highly (up to 70\%) polarized if the local magnetic field is ordered \citep[e.g.][]{bjorbl82}. This polarization signature is commonly detected (at levels up to tens of per cent) in the radio regime from discrete, optically thin jet ejections \citep[e.g.][]{hannet00} and finally confirmed observationally in 2008 at OIR frequencies. One BHXB (GRO J1655--40) and two neutron star XBs (Sco X--1 and Cyg X--2) have linearly polarized NIR emission which is stronger at lower frequencies \citep{shahet08,russfe08}. The polarization levels, $\sim 1$--7\% suggest a twisted magnetic field geometry, but the polarization is probably masked by non-polarized light from other spectral components. The polarization of Sco X--1 is variable on timescales of minutes, which could be explained by a variable flux level of the jet or by a changing magnetic field geometry. The polarization angles imply a magnetic field orientation which differs from source to source and may change in time for the same source \citep{shahet08,russfe08}. Two further BHXBs, XTE J1118+480 and XTE J1550--564, have low levels of polarization ($\sim 0.5$\% in the optical and 1--2\% in the NIR, respectively) which require follow-up observations to confirm their origin \citep{schuet04,dubuch06}.
\newline\newline
\textbf{Baryonic jets:}

The remarkable precessing jets of SS 433 (of which the compact object may be a BH or a neutron star) have signatures at radio, optical and X-ray frequencies. As well as being resolved in the radio and X-ray regimes \citep[e.g.][]{schiet04,miglet05}, broad optical emission lines with moving velocity shifts up to 40,000 km s$^{-1}$ are shown to be from discrete jets moving ballistically at the same velocity as the radio jets: 26\% of the speed of light, at distances $< 10^{10}$ km from their launch region \citep[e.g.][]{murdet80,marget84,vermet93}. This is the only evidence for baryonic relativistic jets from an X-ray binary. Ionized atoms in the jets are  recombining as they cool, and X-ray spectra reveal numerous lines from highly ionized atoms in the jets \citep{marset02} -- their chemical composition therefore includes heavy elements and not pure pair plasma.

SS 433 is surrounded by a large spherical supernova remnant W50. The compact object in SS 433 is thought to be the remnant of the supernova explosion which resulted in W50. The supernova remnant is 10,000 years old. It could be that the recent supernova enriched the photosphere of the companion star with heavy elements, which is now being accreted towards the compact object and swept up into the jets. If this is so, we may not expect to see heavy elements in the jets of other XBs because they are much older, and the photospheres of the companion stars are no longer enriched with heavy elements from the supernova explosion.

A further BHXB that has high-velocity, baryonic outflows is V4641 Sgr. Here, the H$\alpha$ optical and Br$\gamma$ NIR emission lines are very broad with blue wings and variable, indicating an outflow with velocity $\sim 2$\% of the speed of light \citep{chatet03b}. This outflow is however not the same as the highly relativistic (much faster) jet in this source.

\subsection{Internal shocks and impacts with the ISM}

The compact jet, as described in previous Sections, is seen \citep[in the case of Cyg X--1;][]{stiret01} to extend $\sim 10^{8-9}$ km away from the BH. This was measured at one particular radio frequency -- at lower frequencies the compact jet may appear longer still, but will also perhaps be fainter. The compact jet makes only a low contribution to the OIR regime at these large distances from the BH. However bright, discrete, moving ejecta are seen at radio and sometimes X-ray frequencies at distances between $\sim 10^{9}$ and $\sim 10^{14}$ km from the BH, which also emit OIR radiation (Fig. 1). This emission is not part of the overlapping synchrotron spectra that build up the compact jet, but comes from isolated `knots' downstream. Two processes could be responsible: a fast jet plasma cloud catching up with and colliding with a slower ejection, or a plasma cloud that plows into a denser region of the ISM. Both processes result in synchrotron emission. Although such large-scale moving jets have never been detected at OIR frequencies, we discuss here that OIR jet emission does exist on these scales, and could be detected with future observations.

The large-scale, moving jets of XTE J1550--564 were detected at radio and X-ray frequencies and were shown to be decelerating, likely due to interaction with the ISM \citep{corbet02}. The radio spectrum and X-ray flux of the same ejection are consistent with a single power-law of spectral index $\alpha = -0.66$, typical of optically thin synchrotron emission. Deep optical observations were performed with the 8-m Very Large Telescope but no optical counterpart of the ejections were found. The derived optical upper limits were only just greater than the interpolation of the synchrotron power-law, suggesting that with a slightly higher signal-to-noise ratio (a deeper image or a larger telescope) the large-scale ejection would be detected. Deep NIR observations may have been more successful; this source suffers a moderate level of foreground extinction, which affects the optical regime moreso than the NIR.

One claim of a resolved NIR jet from a BHXB does exist in the literature -- that of GRS 1915+105 \citep{samset96}. The jet here is not resolved itself, but appears as a residual to the south-west of the point-spread-function of GRS 1915+105 in the K-band. The implied distance of the jet from the BH is $\sim 6 \times 10^{11}$ km. This could be the NIR counterpart of a discrete ejection with a broadband radio-to-X-ray SED like the one of XTE J1550--564 above.

Extended radio jets (either moving knots or apparently stationary `lobes') are now detected from $\sim 10$ BHXBs. Here, we investigate whether these structures may be observable at OIR frequencies, adopting the assumption that the radio emission is optically thin (observations indicate this is the case), with $\alpha = -0.7$ extrapolated to OIR. We calculate the apparent K-band (2.2$\mu$m) magnitude of the knots given the known foreground extinction to each source. The results are given in Table 1. We see that some of the predicted jet magnitudes are bright enough to be detected with current large telescopes. The brightest, at K $\sim 18$ mag is however only 0.7 arc-seconds from the core, so high-resolution imaging is required, but possible.

\begin{table}
\begin{center}
\small
\caption{The predicted NIR K-band magnitudes of known extended radio jets of BHXBs, assuming $\alpha = -0.7$ from radio to NIR. $\theta$ is the angular distance in arc-seconds of the radio knot from the core. The 2.2$\mu$m values of $F_{\nu}$ are predicted intrinsic flux densities and K mags are the corresponding apparent K-band magnitudes after interstellar extinction is taken into account. For foreground extinction we adopt $A_{\rm K} = 0.114 A_{\rm V}$; estimated values of $A_{\rm V}$ are from \cite{jonket04}, \cite{chapco04} and \cite{eikeet01}.}
\label{tab1}
\begin{tabular}{|llllllll|}
\hline
BHXB&\multicolumn{4}{c}{-------------------- Radio jet knot --------------------}&\multicolumn{2}{c}{-- 2.2$\mu$m flux --}&$A_{\rm V}$\\
    &$F_{\nu}$&Freq.&$\theta$&Reference&$F_{\nu}$&K mag&\\
    &(mJy)&(GHz)&     &         &($\mu$Jy)&&\\
\hline
&\multicolumn{5}{c}{\emph{Moving jets}:}&&\\
XTE J1550--564&3.6&4.8&29''&\cite{corbet02}&2.7&21.6&5.0\\
GRO J1655--40 &120&2.3&0.73''&\cite{tinget95}&55&18.1&3.7\\
GX 339--4     &0.47&4.8&6.9''&\cite{gallet04}&0.36&23.6&3.9\\
V4641 Sgr     &10&4.9&0.6''&\cite{hjelet00}&7.7&19.9&1.0\\
GRS 1915+105  &2.1&5.0&0.23''&\cite{millet05}&1.6&23.8&20\\
\hline
&\multicolumn{5}{c}{\emph{Large-scale lobes}:}&&\\
1E 1740.7--2942&2&1.5&45''&\cite{miraet92}&0.68&27.1&40\\
GRS 1758--258  &0.52&5.0&23''&\cite{rodret92}&0.41&23.9&7.9\\
\hline
\end{tabular}
\normalsize
\end{center}
\end{table}

In addition, the angular size of the resolved X-ray jets of SS 433 are a few arcseconds \citep{miglet05}, so it may be possible to detect the OIR counterparts to these jets with deep observations, possibly in line emission such as shifted H$\alpha$.

SS 433 actually also provides the best evidence yet from the optical regime of a circumbinary disc surrounding an XB -- a separate, non-outflowing large-scale structure around the binary \citep{blunet08}. This result is implied from two stationary H$\alpha$ emission lines; one redshifted and one blueshifted, that suggest an orbiting velocity of $\sim 200$ km s$^{-1}$.

\subsection{Bow shocks, filaments and trails}

At the `front' of the jet there should, in analogy with AGN jets, be a hot spot where the jet slams into dense matter and is decelerated, and a bow shock shell of compressed, shocked gas that surrounds synchrotron-emitting radio plasma \citep[e.g.][an alternative is a gradual deceleration of the jet over large distances by a low-density ISM]{kaiset04}. The shell emits bremsstrahlung radiation and recombination lines. The first confirmed bow shock shell associated with the jet of a BHXB was recently discovered as a bremsstrahlung-emitting radio ring and a recombination line-emitting optical ring, $\sim$10 parsecs away from Cyg X--1 in the direction of its known radio jet \citep{gallet05,russet07a}. This (see Fig. 4) is the largest known optically-emitting structure powered by the jet of a BHXB.

\begin{figure}
\centering
\includegraphics[width=10cm,angle=0]{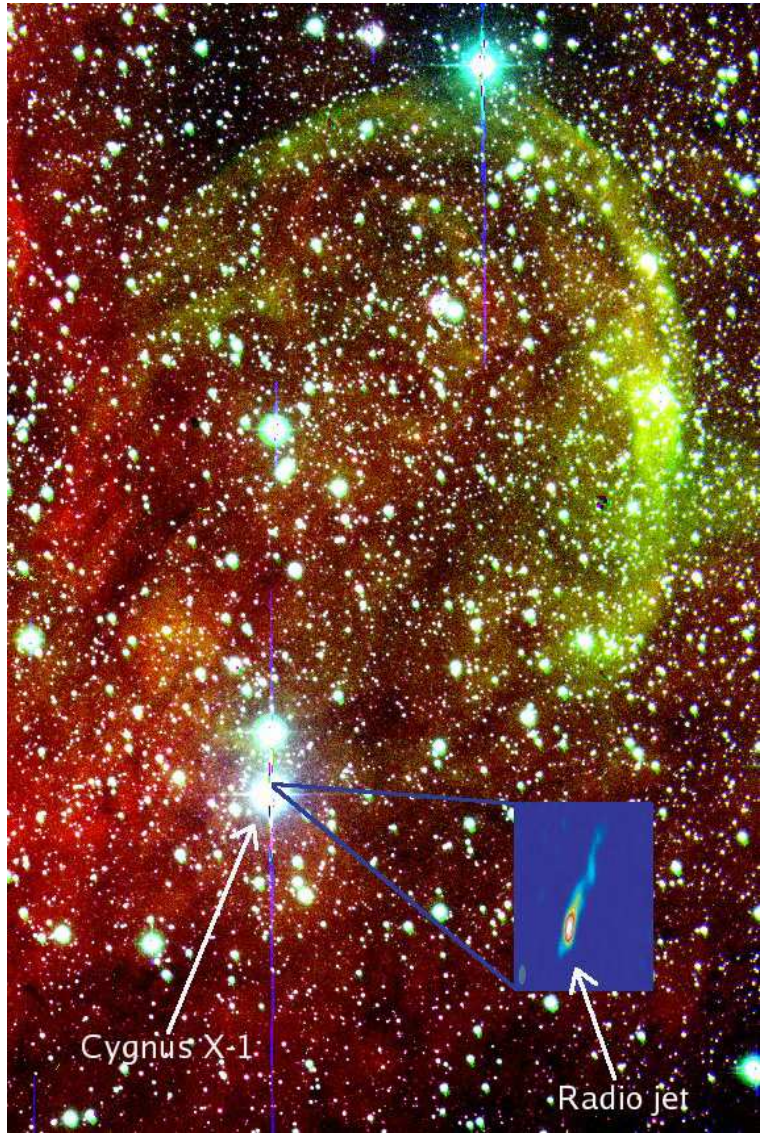}
\caption{Image of the Cyg X--1 ring nebula in optical line emission \citep{russet07a}. The bow shock is powered by the jet of Cyg X--1. H$\alpha$ emission is shown in red, [O III] in green and continuum emission in blue. The nebula appears yellow since the [O III]/H$\alpha$ flux ratio is $\sim 1$, indicating a bow shock velocity $> 100$ km s$^{-1}$. The edge of an H II region can be seen red in the lower left of the image. The compact radio jet of Cyg X--1 (lower right) is $\sim 10^{6}$ times smaller in apparent length \citep[][see Fig. 1]{stiret01}.
}
\label{fig:fig4}
\end{figure}

\cite{cooket07} showed that optical emission line ratios of the nebula close to the BHXB LMC X--1, which was previously thought to be photoionized by the X-rays of the BHXB \citep{pakuan86} revealed the nebula is in some regions shock-excited, and argued this shock excitation arises from the jets of LMC X--1. This is then the second BHXB jet-powered bow shock nebula.

Filaments of optical line emission are also seen where the two jets of SS 433 have broken through the W50 supernova remnant that surround the XB \citep[e.g.][]{boumet07}. These filaments are shock-excited, compressed gas, energized by the jets of SS 433. Rather than a bow shock morphology, the filaments are instead localised regions of shock waves which may contain reflected or secondary shocks. Similar filaments are observed inside the large bow shock nebula of Cyg X--1. It is interesting to note that the emission line ratios of the Cyg X--1 nebula, LMC X--1 nebula and SS 433 filaments all imply shock waves with velocities around $\sim 100$ km s$^{-1}$.

It was argued \citep{hein02} that a high density region of the ISM is needed for a jet-powered bow shock to form and appear bright. XBs with high-mass companions typically occur in denser regions of the ISM than those with low-mass companions. The fact that the three above jet--ISM interaction sites are associated with the jets of XBs with high-mass companions supports this hypothesis.

In addition, it was recently demonstrated that most XBs with low-mass companions are probably travelling through the ISM at velocities greater than a required limit for jet-powered bow shock nebulae to form and remain powered by the jet \citep{heinet08}.
In these cases, synchrotron plasma generated by the jets may form trails behind the XB, if the XB velocity exceeds that of the expanding plasma bubble. These expanding trails will have a shock front where they interact with the ISM. This shocked gas may have been recently detected for a low-mass neutron star XB SAX J1712.6--3739 \citep{wieret09}. Here, two approximately straight stripes of H$\alpha$ emission appear to originate from the location of the XB, analogous to the trails envisaged by \cite{heinet08}. Follow-up imaging and spectroscopy will constrain the properties and nature of this nebulosity.

Nebulae discovered near the XBs GRO J1655--40, LS 5039 and GRS 1009--45 are candidate jet-powered bow shocks and require follow-up observations to confirm these associations \citep{russ08,millet08}.

\section{The jet properties as revealed from OIR analyses}

In Section 2 it is established that the jets of BHXBs can be detected directly and indirectly via numerous methods from optical or infrared observations. We now discuss how these studies can help to constrain the properties of the jets. Due to simple scaling laws that exist theoretically and empirically between the small BHs in XBs and the massive BHs in AGN \citep[Section 1.2; see][for a review]{fend09}, it is possible to infer the properties of AGN jets by studying BHXB jets.

From OIR observations of the compact jet, we can essentially uncover its OIR spectrum (which is usually optically thin; $\alpha \sim -0.7$), variability properties and polarization level/orientation. The OIR flux and spectral index of the compact jet, when combined with the spectrum at lower and higher frequencies, reveals the broadband spectrum of the jet which can then be compared to the most advanced jet models to infer its properties; for example its dimensions, Lorentz factor, confinement, and the total jet power. So far, it has been estimated in some cases that the jet power exceeds the radiative power (which can be derived from the bolometric luminosity). A further dimension of information can constrain the properties more tightly: time-dependent jet modelling. In particular, changes in the jet properties as a function of mass accretion rate and X-ray state will probe the jet physics. Scaling relations with accretion rate and black hole mass can be extrapolated to BHs in AGN, providing predictions that can be tested with observations. If consistent, it could be shown that all BH jets are ubiquitous.

The OIR jet emission of BHXBs originates near the jet base where it is launched and accelerated, so fast timing and polarization studies could provide a wealth of information about the physical conditions here (for example the time-dependent magnetic field configuration or the accretion energy extraction mechanism), providing constraints for jet formation theories. A separate piece of information about the compact jets comes from the moving emission lines from precessing jets when they are seen. These lines are signatures of `heavy' (atomic rather than leptonic) jets and they can also be used to tightly constrain the velocity of the jets.

Extended, moving jet knots detected at OIR frequencies would help to constrain the jet energy losses at large distances downstream. Measurements of deceleration coupled with information about the ISM may tell us how receptive jets are to producing radiation when colliding with dense matter.

Finally, OIR emission lines from ring bow shocks and trailed bow shocks can be used as independent methods of calculating the time-averaged jet power and their composition \citep{gallet05,hein06}. For BHXBs in which this is known and the broadband spectrum of the compact jet has also been measured, the radiative efficiency of the core jet can be derived. Again, these properties can be scaled up to AGN jets. Essentially, how BHXB jets interact with the ISM has direct implications for how AGN jets energize the intracluster medium, shaping galaxy formation.

\section{Conclusion}

To summarise, due to the rapid observational and theoretical advances in jets from BHXBs in the last decade or so, including the finding of direct links with AGN jets, this may be the `golden era' of our understanding of the fastest objects in the universe -- relativistic jets -- and how they are produced via the process of accretion onto compact objects. Perhaps a crucial and satisfactory future goal would be to unify the properties of jets from \emph{all} accreting objects, including the slowest, weakest jets (for example from young stellar objects) and the fastest, most powerful jets (most likely associated with $\gamma$-ray bursts).

From an OIR observer's perspective, there are a number of short term studies in the field of BHXB jets that will help lead us forward in our understanding towards the long term goals:

\begin{itemize}
\item OIR polarization and fast timing analyses of the isolated optically thin jet in many BHXBs, and in different X-ray states when the jet is not bright, to constrain the ubiquity of the properties of the inner regions of the jet;
\item Track the position of the jet break in the spectrum, with changes in mass accretion rate, which constrains the jet power and jet production theories;
\item Resolve the OIR counterparts of discrete, moving radio ejecta;
\item Discover more jet-powered bow shock nebulae and infer the time-averaged jet power for a sample of BHXBs, again testing ubiquity;
\item Compare the properties of BHXB jet-powered structures to AGN lobes and bow shocks, and also to OIR-emitting shock-excited bubbles associated with ultraluminous X-ray sources (which could be intermediate-mass BHs), of which there are many \citep[e.g.][]{pakugr08}.
\item Eventually, it may be possible to resolve the compact jet at OIR frequencies indirectly using interferometry \citep{mark08}, which will open up new avenues of studying the inner regions of the jets and accretion flow.
\end{itemize}

\vspace{5mm}
\emph{Acknowledgements}.
DMR would like to thank Piergiorgio Casella for useful discussions regarding the interpretations of multi-wavelength fast-timing analyses. DMR acknowledges support from a Netherlands Organization for Scientific Research (NWO) Veni Fellowship.

\label{lastpage-01}


\begin{thebibliography}{natbib}
\bibitem[\protect\citeauthoryear{Baade \& Minkowski}{1954}]{baadmi54}Baade, W.; Minkowski, R. \emph{ApJ} 1954, 119, 221
\bibitem[\protect\citeauthoryear{Bartolini et al.}{1994}]{bartet94}Bartolini, C.; Guarnieri, A.; Piccioni, A.; Beskin, G. M.; Neizvestny, S. I. \emph{ApJS} 1994, 92, 455
\bibitem[\protect\citeauthoryear{Bj\"ornsson \& Blumenthal}{1982}]{bjorbl82}Bj\"ornsson, C.-I.; Blumenthal, G. R. \emph{ApJ} 1982, 259, 805
\bibitem[\protect\citeauthoryear{Blandford \& Znajek}{1977}]{blanzn77}Blandford, R. D.; Znajek, R. L. \emph{MNRAS} 1977, 179, 433
\bibitem[\protect\citeauthoryear{Blandford \& Konigl}{1979}]{blanko79}Blandford, R. D.; Konigl, A. \emph{ApJ} 1979, 232, 34
\bibitem[\protect\citeauthoryear{Blandford \& Payne}{1982}]{blanpa82}Blandford, R. D.; Payne, D. G. \emph{MNRAS} 1982, 199, 883
\bibitem[\protect\citeauthoryear{Blundell et al.}{2008}]{blunet08}Blundell, K. M.; Bowler, M. G.; Schmidtobreick, L. \emph{ApJ} 2008, 678, L47
\bibitem[\protect\citeauthoryear{Boumis et al.}{2007}]{boumet07}Boumis, P.; et al. \emph{MNRAS} 2007, 381, 308
\bibitem[\protect\citeauthoryear{Brocksopp et al.}{2001}]{brocet01}Brocksopp, C.; Jonker, P. G.; Fender, R. P.; Groot, P. J.; van der Klis, M.; Tingay, S. J. \emph{MNRAS} 2001, 323, 517
\bibitem[\protect\citeauthoryear{Brocksopp, Bandyopadhyay \& Fender}{Brocksopp et al.}{2004}]{brocet04}Brocksopp, C.; Bandyopadhyay, R. M.; Fender, R. P. \emph{NewA} 2004, 9, 249
\bibitem[\protect\citeauthoryear{Burbidge}{1959}]{burb59}Burbidge, G. R. \emph{ApJ} 1959, 129, 849
\bibitem[\protect\citeauthoryear{Buxton \& Bailyn}{2004}]{buxtet04}Buxton, M. M.; Bailyn, C. D. \emph{ApJ} 2004, 615, 880
\bibitem[\protect\citeauthoryear{Chakrabarti et al.}{2005}]{chaket05}Chakrabarti, S. K.; et al. \emph{MNRAS} 2005, 362, 957
\bibitem[\protect\citeauthoryear{Chapuis \& Corbel}{2004}]{chapco04}Chapuis, C.; Corbel, S. \emph{A\&A} 2004, 414, 659
\bibitem[\protect\citeauthoryear{Chaty et al.}{2003a}]{chatet03a}Chaty, S.; Haswell, C. A.; Malzac, J.; Hynes, R. I.; Shrader, C. R.; Cui W. \emph{MNRAS} 2003a, 346, 689
\bibitem[\protect\citeauthoryear{Chaty et al.}{2003b}]{chatet03b}Chaty, S.; Charles, P. A.; Mart\'i, J.; Mirabel, I. F.; Rodr\'iguez, L. F.; Shahbaz, T. \emph{MNRAS} 2003b, 343, 169
\bibitem[\protect\citeauthoryear{Churazov et al.}{2002}]{churet02}Churazov, E.; Sunyaev, R.; Forman, W.; B\"ohringer, H. \emph{MNRAS} 2002, 332, 729
\bibitem[\protect\citeauthoryear{Cooke et al.}{2007}]{cooket07}Cooke, R.; Kuncic, Z.; Sharp, R.; Bland-Hawthorn, J. \emph{ApJL} 2007, 667, 163
\bibitem[\protect\citeauthoryear{Corbel \& Fender}{2002}]{corbfe02}Corbel, S.; Fender, R. P. \emph{ApJ} 2002, 573, L35
\bibitem[\protect\citeauthoryear{Corbel et al.}{2002}]{corbet02}Corbel, S.; Fender, R. P.; Tzioumis, A. K.; Tomsick, J. A.; Orosz, J. A.; Miller, J. M.; Wijnands, R.; Kaaret, P. \emph{Science} 2002, 298, 196
\bibitem[\protect\citeauthoryear{Dubus \& Chaty}{2006}]{dubuch06}Dubus, G.; Chaty, S. \emph{A\&A} 2006, 458, 591
\bibitem[\protect\citeauthoryear{Durant et al.}{2008}]{duraet08}Durant, M.; Gandhi, P.; Shahbaz, T.; Fabian, A. P.; Miller, J.; Dhillon, V. S.; Marsh, T. R. \emph{ApJ} 2008, 682, L45
\bibitem[\protect\citeauthoryear{Durant et al.}{2009}]{duraet09}Durant, M.; Gandhi, P.; Shahbaz, T.; Peralta, H. H.; Dhillon, V. S. \emph{MNRAS} 2009, 392, 309
\bibitem[\protect\citeauthoryear{Efstathiou}{1992}]{efst92}Efstathiou, G. \emph{MNRAS} 1992, 256, P43
\bibitem[\protect\citeauthoryear{Eikenberry et al.}{1998a}]{eikeet98a}Eikenberry, S. S.; Matthews, K.; Murphy, T. W. Jr.; Nelson, R. W.; Morgan, E. H.; Remillard, R. A.; Muno, M. \emph{ApJ} 1998a, 506, L31
\bibitem[\protect\citeauthoryear{Eikenberry et al.}{1998b}]{eikeet98b}Eikenberry, S. S.; Matthews, K.; Morgan, E. H.; Remillard, R. A.; Nelson, R. W. \emph{ApJ} 1998b, 494, L61
\bibitem[\protect\citeauthoryear{Eikenberry et al.}{2001}]{eikeet01}Eikenberry, S. S.; Fischer, W. J.; Egami, E.; Djorgovski, S. G. \emph{ApJ} 2001, 556, 1
\bibitem[\protect\citeauthoryear{Fabian et al.}{1982}]{fabiet82}Fabian, A. C.; Guilbert, P. W.; Motch, C.; Ricketts, M.; Ilovaisky, S. A.; Chevalier, C. \emph{A\&A} 1982, 111, L9
\bibitem[\protect\citeauthoryear{Falcke \& Biermann}{1996}]{falcbi96}Falcke, H.; Biermann, P. L. \emph{A\&A} 1996, 308, 321
\bibitem[\protect\citeauthoryear{Falcke, K\"ording \& Markoff}{Falcke et al.}{2004}]{falcet04}Falcke, H.; K\"ording, E.; Markoff, S. \emph{A\&A} 2004, 414, 895
\bibitem[\protect\citeauthoryear{Fanaroff \& Riley}{1974}]{fanari74}Fanaroff, B. L.; Riley, J. M. \emph{MNRAS} 1974, 167, P31
\bibitem[\protect\citeauthoryear{Fender}{2001}]{fend01}Fender, R. P. \emph{MNRAS} 2001, 322, 31
\bibitem[\protect\citeauthoryear{Fender}{2006}]{fend06}Fender, R. P. in \emph{Compact Stellar X-Ray Sources}; Cambridge University Press: Cambridge, U.K., 2006; pp 381 (eds. Lewin, W. H. G.; van der Klis, M.)
\bibitem[\protect\citeauthoryear{Fender}{2009}]{fend09}Fender, R. P. \emph{Disc-jet coupling in black hole X-ray binaries and active galactic nuclei}; in The Jet Paradigm - From Microquasars to Quasars, Lect. Notes Phys. 794 (ed. Belloni, T.) (arXiv:0909.2572)
\bibitem[\protect\citeauthoryear{Fender et al.}{1997}]{fendet97}Fender, R. P.; Pooley, G. G.; Brocksopp, C.; Newell, S. J. \emph{MNRAS} 1997, 290, L65
\bibitem[\protect\citeauthoryear{Fender \& Pooley}{1998}]{fendpo98}Fender, R. P.; Pooley, G. G. \emph{MNRAS} 1998, 300, 573
\bibitem[\protect\citeauthoryear{Fender \& Pooley}{2000}]{fendpo00}Fender, R. P.; Pooley, G. G. \emph{MNRAS} 2000, 318, L1
\bibitem[\protect\citeauthoryear{Fender et al.}{2001}]{fendet01}Fender, R. P.; Hjellming, R. M.; Tilanus, R. P. J.; Pooley, G. G.; Deane, J. R.; Ogley, R. N.; Spencer, R. E. \emph{MNRAS} 2001, 322, L23
\bibitem[\protect\citeauthoryear{Fender, Belloni \& Gallo}{Fender et al.}{2004}]{fendet04}Fender, R. P.; Belloni, T. M.; Gallo, E. \emph{MNRAS} 2004, 355, 1105
\bibitem[\protect\citeauthoryear{Fender et al.}{2009}]{fendet09}Fender, R. P.; Russell, D. M.; Knigge, C.; Soria, R.; Hynes, R. I.; Goad, M. \emph{MNRAS} 2009, 393, 1608
\bibitem[\protect\citeauthoryear{Flanagan \& Hughes}{1998}]{flanet98}Flanagan, \'E. \'E.; Hughes, S. A.; \emph{PhRvD} 1998, 57, 4535
\bibitem[\protect\citeauthoryear{Fomalont, Geldzahler \& Bradshaw}{Fomalont et al.}{2001}]{fomaet01}Fomalont, E. B.; Geldzahler, B. J.; Bradshaw, C. F. \emph{ApJ} 2001, 558, 283
\bibitem[\protect\citeauthoryear{Gallo, Fender \& Pooley}{Gallo et al.}{2003}]{gallet03}Gallo, E.; Fender, R. P.; Pooley, G. G. \emph{MNRAS} 2003, 344, 60
\bibitem[\protect\citeauthoryear{Gallo et al.}{2004}]{gallet04}Gallo, E.; Corbel, S.; Fender, R. P.; Maccarone, T. J.; Tzioumis, A. K. \emph{MNRAS} 2004, 347, L52
\bibitem[\protect\citeauthoryear{Gallo et al.}{2005}]{gallet05}Gallo, E.; Fender, R. P.; Kaiser, C.; Russell, D. M.; Morganti, R.; Oosterloo, T.; Heinz, S. \emph{Nature} 2005, 436, 819
\bibitem[\protect\citeauthoryear{Gallo et al.}{2006}]{gallet06}Gallo, E.; et al. \emph{MNRAS} 2006, 370, 1351
\bibitem[\protect\citeauthoryear{Gallo et al.}{2007}]{gallet07}Gallo, E.; Migliari, S.; Markoff, S.; Tomsick, J.; Bailyn, C.; Berta, S.; Fender, R.; Miller-Jones, J. \emph{ApJ} 2007, 670, 600
\bibitem[\protect\citeauthoryear{Gandhi et al.}{2008}]{gandet08}Gandhi, P.; et al. \emph{MNRAS} 2008, 390, L29
\bibitem[\protect\citeauthoryear{Goranskii et al.}{1998}]{goraet98}Goranskii, V. P.; Esipov, V. F.; Cherepashchuk, A. M. \emph{ARep} 1998, 42, 336
\bibitem[\protect\citeauthoryear{Han \& Hjellming}{1992}]{hanet92}Han, X.; Hjellming, R. M. \emph{ApJ} 1992, 400, 304
\bibitem[\protect\citeauthoryear{Hannikainen et al.}{2000}]{hannet00}Hannikainen, D. C.; Hunstead, R. W.; Campbell-Wilson, D.; Wu, K.; McKay, D. J.; Smits, D. P.; Sault, R. J. \emph{ApJ} 2000, 540, 521
\bibitem[\protect\citeauthoryear{Heinz}{2002}]{hein02}Heinz, S. \emph{A\&A} 2002, 388, L40
\bibitem[\protect\citeauthoryear{Heinz}{2006}]{hein06}Heinz, S. \emph{ApJ} 2006, 636, 316
\bibitem[\protect\citeauthoryear{Heinz \& Sunyaev}{2003}]{heinsu03}Heinz, S.; Sunyaev, R. A. \emph{MNRAS} 2003, 343, L59
\bibitem[\protect\citeauthoryear{Heinz et al.}{2008}]{heinet08}Heinz, S.; Grimm, H. J.; Sunyaev, R. A.; Fender, R. P. \emph{ApJ} 2008, 686, 1145
\bibitem[\protect\citeauthoryear{Hjellming et al.}{2000}]{hjelet00}Hjellming, R. M.; et al. \emph{ApJ} 2000, 544, 977
\bibitem[\protect\citeauthoryear{Homan et al.}{2005}]{homaet05}Homan, J.; Buxton, M.; Markoff, S.; Bailyn, C. D.; Nespoli, E.; Belloni, T. \emph{ApJ} 2005, 624, 295
\bibitem[\protect\citeauthoryear{Hynes}{2005}]{hyne05}Hynes, R. I. \emph{ApJ} 2005, 623, 1026
\bibitem[\protect\citeauthoryear{Hynes et al.}{1998}]{hyneet98}Hynes, R. I.; O'Brien, K.; Horne, K.; Chen, W.; Haswell, C. A. \emph{MNRAS} 1998, 299, L37
\bibitem[\protect\citeauthoryear{Hynes \& Haswell}{1999}]{hyneha99}Hynes, R. I.; Haswell, C. A. \emph{MNRAS} 1999, 303, 101
\bibitem[\protect\citeauthoryear{Hynes et al.}{2000}]{hyneet00}Hynes, R. I.; Mauche, C. W.; Haswell, C. A.; Shrader, C. R.; Cui, W.; Chaty, S. \emph{ApJ} 2000, 539, L37
\bibitem[\protect\citeauthoryear{Hynes et al.}{2002}]{hyneet02}Hynes, R. I.; Haswell, C. A.; Chaty, S.; Shrader, C. R.; Cui, W. \emph{MNRAS} 2002, 331, 169
\bibitem[\protect\citeauthoryear{Hynes et al.}{2003a}]{hyneet03a}Hynes, R. I.; Charles, P. A.; Casares, J.; Haswell, C. A.; Zurita, C.; Shahbaz, T. \emph{MNRAS} 2003a, 340, 447
\bibitem[\protect\citeauthoryear{Hynes et al.}{2003b}]{hyneet03b}Hynes, R. I.; et al. \emph{MNRAS} 2003b, 345, 292
\bibitem[\protect\citeauthoryear{Hynes et al.}{2004}]{hyneet04}Hynes, R. I.; et al. \emph{ApJ} 2004, 611, L125
\bibitem[\protect\citeauthoryear{Hynes et al.}{2006}]{hyneet06}Hynes, R. I.; et al. \emph{ApJ} 2006, 651, 401
\bibitem[\protect\citeauthoryear{Jain et al.}{2001}]{jainet01}Jain, R. K.; Bailyn, C. D.; Orosz, J. A.; McClintock, J. E.; Remillard, R. A. \emph{ApJ} 2001, 554, L181
\bibitem[\protect\citeauthoryear{Jamil et al.}{2008}]{jamiet08}Jamil, O.; Fender, R.; Kaiser, C. \emph{Proceedings of Science}, 2008, \emph{Proceedings of the VII Microquasar Workshop: Microquasars and Beyond, 1-5 September 2008, Foca, Izmir, Turkey} (arXiv:0811.3320)
\bibitem[\protect\citeauthoryear{Jonker \& Nelemans}{2004}]{jonket04}Jonker, P. G.; Nelemans, G. \emph{MNRAS} 2004, 354, 355
\bibitem[\protect\citeauthoryear{Kaiser}{2006}]{kais06}Kaiser, C. R. \emph{MNRAS} 2006, 367, 1083
\bibitem[\protect\citeauthoryear{Kaiser et al.}{2004}]{kaiset04}Kaiser, C. R.; Gunn, K. F.; Brocksopp, C.; Sokoloski, J. L. \emph{ApJ} 2004, 612, 332
\bibitem[\protect\citeauthoryear{Kalemci et al.}{2005}]{kaleet05}Kalemci, E.; et al. \emph{ApJ} 2005, 622, 508
\bibitem[\protect\citeauthoryear{Kanbach et al.}{2001}]{kanbet01}Kanbach, G.; Straubmeier, C.; Spruit, H. C.; Belloni, T. \emph{Nature} 2001, 414, 180
\bibitem[\protect\citeauthoryear{Klein-Wolt et al.}{2002}]{kleiet02}Klein-Wolt, M.; Fender, R. P.; Pooley, G. G.;p Belloni, T.; Migliari, S.; Morgan, E. H.; van der Klis, M. \emph{MNRAS} 2002, 331, 745
\bibitem[\protect\citeauthoryear{K\"ording, Fender \& Migliari}{K\"ording et al.}{2006a}]{kordet06a}K\"ording, E.; Fender, R. P.; Migliari, S. \emph{MNRAS} 2006a, 369, 1451
\bibitem[\protect\citeauthoryear{K\"ording, Jester \& Fender}{K\"ording et al.}{2006b}]{kordje06}K\"ording, E. G.; Jester, S.; Fender, R. \emph{MNRAS} 2006b, 372, 1366
\bibitem[\protect\citeauthoryear{Kuulkers}{1998}]{kuul98}Kuulkers, E. \emph{NewAR} 1998, 42, 1
\bibitem[\protect\citeauthoryear{Maitra \& Bailyn}{2008}]{maitba08}Maitra, D.; Bailyn, C. D. \emph{ApJ} 2008, 688, 537
\bibitem[\protect\citeauthoryear{Malzac, Merloni \& Fabian}{Malzac et al.}{2004}]{malzet04}Malzac, J.; Merloni, A.; Fabian, A. C. \emph{MNRAS} 2004, 351, 253
\bibitem[\protect\citeauthoryear{Margon et al.}{1984}]{marget84}Margon, B.; Anderson, S. F.; Aller, L. H.; Downes, R. A.; Keyes, C. D. \emph{ApJ} 1984, 281, 313
\bibitem[\protect\citeauthoryear{Markoff}{2008}]{mark08}Markoff, S. \emph{Proceedings of Science}, 2008, \emph{Proceedings of the VII Microquasar Workshop: Microquasars and Beyond, 1-5 September 2008, Foca, Izmir, Turkey} (arXiv:0811.3601)
\bibitem[\protect\citeauthoryear{Markoff, Falcke \& Fender}{Markoff et al.}{2001}]{market01}Markoff, S.; Falcke, H.; Fender, R. \emph{A\&A} 2001, 372, L25
\bibitem[\protect\citeauthoryear{Markoff et al.}{2003}]{market03}Markoff, S.; Nowak, M.; Corbel, S.; Fender, R.; Falcke, H. \emph{A\&A} 2003, 397, 645
\bibitem[\protect\citeauthoryear{Markoff, Nowak \& Wilms}{Markoff et al.}{2005}]{market05}Markoff, S.; Nowak, M. A.; Wilms, J. \emph{ApJ} 2005, 635, 1203
\bibitem[\protect\citeauthoryear{Marshall et al.}{2002}]{marset02}Marshall, H. L.; Canizares, C. R.; Schulz, N. S. \emph{ApJ} 2002, 564, 941
\bibitem[\protect\citeauthoryear{Massey, Johnson \& Degioia-Eastwood}{Massey et al.}{1995}]{masset95}Massey, P.; Johnson, K. E.; Degioia-Eastwood, K. \emph{ApJ} 1995, 454, 151
\bibitem[\protect\citeauthoryear{McNamara \& Nulsen}{2007}]{mcnaet07}McNamara, B. R.; Nulsen, P. E. J. \emph{ARA\&A} 2007, 45, 117
\bibitem[\protect\citeauthoryear{Meier}{2001}]{meie01}Meier, D. L. \emph{ApJ} 2001, 548, L9
\bibitem[\protect\citeauthoryear{Merloni, Heinz \& di Matteo}{Merloni et al.}{2003}]{merlet03}Merloni, A.; Heinz, S.; di Matteo, T. \emph{MNRAS} 2003, 345, 1057
\bibitem[\protect\citeauthoryear{Migliari et al.}{2005}]{miglet05}Migliari, S.; Fender, R. P.; Blundell, K. M.; M\'endez, M.; van der Klis, M. \emph{MNRAS} 2005, 358, 860
\bibitem[\protect\citeauthoryear{Migliari \& Fender}{2006}]{miglfe06}Migliari, S.; Fender, R. P. \emph{MNRAS} 2006, 366, 79
\bibitem[\protect\citeauthoryear{Migliari et al.}{2007}]{miglet07}Migliari, S.; et al. \emph{ApJ} 2007, 670, 610
\bibitem[\protect\citeauthoryear{Miller-Jones et al.}{2005}]{millet05}Miller-Jones, J. C. A.; McCormick, D. G.; Fender, R. P.; Spencer, R. E.; Muxlow, T. W. B.; Pooley, G. G. \emph{MNRAS} 2005, 363, 867
\bibitem[\protect\citeauthoryear{Miller-Jones et al.}{2006}]{millet06}Miller-Jones, J. C. A.; Fender, R. P.;  Nakar, E. \emph{MNRAS} 2006, 367, 1432
\bibitem[\protect\citeauthoryear{Miller-Jones et al.}{2008}]{millet08}Miller-Jones, J.; Russell, D.; Brocksopp, C.; Sokoloski, J.; Stappers, B.; Muxlow, T. \emph{AIPC} 2008, 1010, 50 (arXiv:0802.3446)
\bibitem[\protect\citeauthoryear{Mirabel et al.}{1992}]{miraet92}Mirabel, I. F.; Rodr\'{i}guez, L. F.; Cordier, B.; Paul, J.; Lebrun, F. \emph{Nature} 1992, 358, 215
\bibitem[\protect\citeauthoryear{Mirabel et al.}{1998}]{miraet98}Mirabel, I. F.; Dhawan, V.; Chaty, S.; Rodriguez, L. F.; Marti, J.; Robinson, C. R.; Swank, J.; Geballe, T. \emph{A\&A} 1998, 330, L9
\bibitem[\protect\citeauthoryear{Motch, Ilovaisky \& Chevalier}{Motch et al.}{1982}]{motcet82}Motch, C.; Ilovaisky, S. A.; Chevalier, C. \emph{A\&A} 1982, 109, L1
\bibitem[\protect\citeauthoryear{Motch et al.}{1983}]{motcet83}Motch, C.; Ricketts, M. J.; Page, C. G.; Ilovaisky, S. A.; Chevalier, C. \emph{A\&A} 1983, 119, 171
\bibitem[\protect\citeauthoryear{Motch et al.}{1985}]{motcet85}Motch, C.; Ilovaisky, S. A.; Chevalier, C.; Angebault, P. \emph{SSRv} 1985, 40, 219
\bibitem[\protect\citeauthoryear{Murdin et al.}{1980}]{murdet80}Murdin, P.; Clark, D. H.; Martin, P. G. \emph{MNRAS} 1980, 193, 135
\bibitem[\protect\citeauthoryear{Nowak et al.}{2005}]{nowaet05}Nowak, M. A.; Wilms, J.; Heinz, S.; Pooley, G.; Pottschmidt, K.; Corbel, S. \emph{ApJ} 2005, 626, 1006
\bibitem[\protect\citeauthoryear{O'Brien et al.}{2002}]{obriet02}O'Brien, K.; Horne, K.; Hynes, R. I.; Chen, W.; Haswell, C. A.; Still, M. D. \emph{MNRAS} 2002, 334, 426
\bibitem[\protect\citeauthoryear{Pakull \& Angebault}{1986}]{pakuan86}Pakull, M. W.; Angebault, L. P. \emph{Nature} 1986, 322, 511
\bibitem[\protect\citeauthoryear{Pakull \& Gris\'e}{2008}]{pakugr08}Pakull, M. W.; Gris\'e, F. \emph{AIPC} 2008, 1010, 303 (arXiv:0803.4345)
\bibitem[\protect\citeauthoryear{Pineault}{1984}]{pine84}Pineault, S. \emph{A\&A} 1984, 139, 313
\bibitem[\protect\citeauthoryear{Rodr\'{i}guez et al.}{1992}]{rodret92}Rodr\'{i}guez, L. F.; Mirabel, I. F.; Mart\'{i}, J. \emph{ApJ} 1992, 401, L15
\bibitem[\protect\citeauthoryear{Rothstein et al.}{2005}]{rothet05}Rothstein, D. M.; Eikenberry, S. S.; Matthews, K. \emph{ApJ} 2005, 626, 991
\bibitem[\protect\citeauthoryear{Russell}{2008}]{russ08}Russell, D. M. \emph{Ph.D. thesis} 2008 (arXiv:0802.0816)
\bibitem[\protect\citeauthoryear{Russell et al.}{2006}]{russet06}Russell, D. M.; Fender, R. P.; Hynes, R. I.; Brocksopp, C.; Homan, J.; Jonker, P. G.; Buxton, M. M. \emph{MNRAS} 2006, 371, 1334
\bibitem[\protect\citeauthoryear{Russell et al.}{2007a}]{russet07a}Russell, D. M.; Fender, R. P.; Gallo, E.; Kaiser, C. R. \emph{MNRAS} 2007a, 376, 1341
\bibitem[\protect\citeauthoryear{Russell et al.}{2007b}]{russet07b}Russell, D. M.; Maccarone, T. J.; K\"ording, E. G.; Homan, J. \emph{MNRAS} 2007b, 379, 1401
\bibitem[\protect\citeauthoryear{Russell \& Fender}{2008}]{russfe08}Russell, D. M.; Fender, R. P. \emph{MNRAS} 2008, 387, 713
\bibitem[\protect\citeauthoryear{Russell et al.}{2008}]{russet08}Russell, D. M.; Maitra, D.; Fender, R. P.; Lewis, F. \emph{Proceedings of Science}, 2008, \emph{Proceedings of the VII Microquasar Workshop: Microquasars and Beyond, 1-5 September 2008, Foca, Izmir, Turkey} (arXiv:0811.2919)
\bibitem[\protect\citeauthoryear{Sams, Eckart \& Sunyaev}{Sams et al.}{1996}]{samset96}Sams, B. J.; Eckart, A.; Sunyaev, R. \emph{Nature} 1996, 382, 47
\bibitem[\protect\citeauthoryear{Schawinski et al.}{2009}]{schaet09}Schawinski, K.; Virani, S.; Simmons, B.; Urry, C. M.; Treister, E.; Kaviraj, S.; Kushkuley, B. \emph{ApJ} 2009, 692, L19
\bibitem[\protect\citeauthoryear{Schillemat et al.}{2004}]{schiet04}Schillemat, K.; Mioduszewski, A.; Dhawan, V.; Rupen, M. \emph{AAS} 2004, 20510401
\bibitem[\protect\citeauthoryear{Schultz, Hakala \& Huovelin}{Schultz et al.}{2004}]{schuet04}Schultz, J.; Hakala, P.; Huovelin, J. \emph{BaltA} 2004, 13, 581
\bibitem[\protect\citeauthoryear{Shahbaz et al.}{2003}]{shahet03}Shahbaz, T.; et al. \emph{MNRAS} 2003, 346, 1116
\bibitem[\protect\citeauthoryear{Shahbaz et al.}{2004}]{shahet04}Shahbaz, T.; et al. \emph{MNRAS} 2004, 354, 31
\bibitem[\protect\citeauthoryear{Shahbaz et al.}{2008}]{shahet08}Shahbaz, T.; Fender, R. P.; Watson, C. A.; O'Brien, K. \emph{ApJ} 2008, 672, 510
\bibitem[\protect\citeauthoryear{Shrader et al.}{1994}]{shraet94}Shrader, C. R.; Wagner, R. M.; Hjellming, R. M.; Han, X. H.; Starrfield, S. G. \emph{ApJ} 1994, 434, 698
\bibitem[\protect\citeauthoryear{Stirling et al.}{2001}]{stiret01}Stirling, A. M.; Spencer, R. E.; de la Force, C. J.; Garrett, M. A.; Fender, R. P.; Ogley, R. N. \emph{MNRAS} 2001, 327, 1273
\bibitem[\protect\citeauthoryear{Tingay et al.}{1995}]{tinget95}Tingay, S. J.; et al. \emph{Nature} 1995, 374, 141
\bibitem[\protect\citeauthoryear{Treves et al.}{1988}]{trevet88}Treves, A.; Belloni, T.; Bouchet, P.; Chiappetti, L.; Falomo, R.; Maraschi, L.; Tanzi, E. G. \emph{ApJ} 1988, 335, 142
\bibitem[\protect\citeauthoryear{Ueda et al.}{2002}]{uedaet02}Ueda, Y.; et al. \emph{ApJ} 2002, 571, 918
\bibitem[\protect\citeauthoryear{Uemura et al.}{2002}]{uemuet02}Uemura, M.; et al. \emph{PASJ} 2002, 54, L79
\bibitem[\protect\citeauthoryear{Uemura et al.}{2004a}]{uemuet04a}Uemura, M.; et al. \emph{PASJ} 2004a, 56, 61
\bibitem[\protect\citeauthoryear{Uemura et al.}{2004b}]{uemuet04b}Uemura, M.; et al. \emph{PASJ} 2004b, 56, 823
\bibitem[\protect\citeauthoryear{van Paradijs \& McClintock}{1994}]{vanpet94}van Paradijs, J.; McClintock, J. E. \emph{A\&A} 1994, 290, 133
\bibitem[\protect\citeauthoryear{Vermeulen et al.}{1993}]{vermet93}Vermeulen, R. C.; et al. \emph{A\&A} 1993, 270, 204
\bibitem[\protect\citeauthoryear{Wiersema et al.}{2009}]{wieret09}Wiersema, K.; Russell, D. M.; Degenaar, N.; Klein-Wolt, M.; Wijnands, R.; Heinz, S.; Read, A. M.; Saxton, R. D.; Tanvir, N. R. \emph{MNRAS} 2009, 397, L6
\bibitem[\protect\citeauthoryear{Zurita, Casares \& Shahbaz}{Zurita et al.}{2003}]{zuriet03}Zurita, C.; Casares, J.; Shahbaz, T. \emph{ApJ} 2003, 582, 369
\end{thebibliography}
\end{document}